# A distribution function correction-based immersed boundary-lattice Boltzmann method with truly second-order accuracy for fluid-solid flows

Shi Tao*, Qing He, Baiman Chen, Simin Huang

*Key Laboratory of Distributed Energy Systems of Guangdong Province, Dongguan University of Technology, Dongguan 523808, China*

**Abstract**

The immersed boundary lattice Boltzmann method (IB-LBM) has been widely used in the simulation of fluid-solid interaction and particulate flow problems, since proposed in 2004. However, it is usually a non-trivial task to retain the flexibility and the accuracy simultaneously in the implementation of this approach. Based on the intrinsic nature of the LBM, we propose a simple and second-order accurate IB-LBM in this paper, where the no-slip boundary condition on the fluid-solid interface is directly imposed by iteratively correcting the distribution function near the boundary, markedly similar to the treatment of boundary condition in the original LBM. Therefore, there are no additional efforts needed to construct an interfacial force model (such as the spring, feedback and direct force models) and absorb this force into the framework of LBM. It is worth mentioning that we retain the discrete delta function for the connection between the Lagrangian and Eulerian meshes, thus preserving the flexibility of the conventional IB-LBM. Second-order accuracy of the present IB-LBM is reasonably confirmed in the simulation of cylindrical Couette flow, while the conventional approach is generally a first-order scheme. The present method is first validated in the flow past a fixed cylinder. No streamline penetration is found, indicating the exactly enforcement of the no-slip boundary condition. Furthermore, in the simulations of one and two particles settling under gravity, good agreements can be observed comparing with the data available in the literature.

## 1. Introduction

The immersed boundary method (IBM) is a very popular numerical approach in the simulation of complex and moving boundary problems [1–3]. Since it is generally based on the fixed Cartesien grids, the time-consuming procedure for mesh-generation and remeshing involved in other methods, such as the Arbitrary-Lagrange-Euler (ALE) that uses a body-fitted grid and regenerates the computation mesh whenever the boundary moves, can be totally removed. The IBM is usually applied to handle the boundary of a body. Therefore, an efficient flow solver is needed to simulate the flow of the carrier fluid. In most cases in the past, IBM was combined with the Navier-Stokes (N-S) solvers. Recently, some efforts were also made to combine IBM with the lattice Boltzmann method (LBM) [4–6]. Compared to the traditional N-S solvers, a noteworthy feature of LBM is that it is a kinetic scheme. Hence, there is no need for LBM to solve the time-consuming Poisson equation, since the pressure in LBM is determined directly by the equation of state. In addition, the LBM has some other distinctive merits, such as easy implementation and natural parallelism. For those reasons, the application of LBM as an alternative flow solver has been advanced dramatically in recent years [7,8]. A pioneering work from Feng et al. [4] first successfully combined the IBM with LBM in 2004, considering that the LBM is also based on the uniform Cartesien grids. This hybrid approach, named IB-LBM, has then been demonstrated to well preserve the mutual advantages of the individual methods, and more and more widely used to deal with the fluid-solid interaction and particulate flow problems [9–12,66].

---

*lssts1013@hust.edu.cn (Shi Tao)

In IBM, the solid boundary is usually modeled as a generator of external force. This force will be added into the governing equation of fluid, and can then affect the macroscopic quantities. So that the no-slip boundary condition on the surface of immersed body is to be imposed. The IBM generally falls into two categories: the continuous forcing and discrete forcing methods [13]. The discrete forcing method is a sharp interface scheme, where the external body force is applied unilaterally around the boundary or contained implicitly in the discrete governing equation. Even though the discrete forcing approach has high accuracy for the boundary treatment, a major disadvantage is that serious spurious-force-oscillation (SFO) is almost inevitably to emerge when the boundary moves [14-16]. It is still a challenging and ongoing subject for researchers to remove such deficiency. Furthermore, the original LBM itself is a sharp interface scheme (the bounce-back type methods used in the boundary treatment), and hence experiences the same problem of SFO when applied it to moving boundaries [17,18,61]. Together with the complex algorithm for application in the discrete forcing method, all those result in that it is not so attractive to combine that scheme with LBM. On the other hand, continuous forcing is a diffuse interface scheme, where the solid interface smears in several grids at both sides of the boundary. The bilateral fluid is treated in the same manner, which drastically reduces the difficulty in applications. An accompanying advantage is that the phenomenon of SFO can be well removed in the continuous forcing method. In general, the focus of the recent studies seems to be more on the development of continuous forcing IB-LBM [10,11,19,20] (the term continuous forcing is omitted hereinafter when referring to IB-LBM).

A tricky problem presently experienced by the IB-LBM is the low precision (usually only first-order accuracy) for the boundary treatment. Some authors considered it may be caused by the streamline penetration, a phenomenon first observed by Shu et al. [21] in the simulation of flow past a fixed circular cylinder. This drawback is subsequently resolved by Wu et al. [5], using a velocity correction- based implicit forcing technology. Note that the multi-forcing strategy can also be used to efficiently remedy the deficiency of streamline penetration [22,20]. Furthermore, it has been founded that those above-mentioned boundary condition-enforced IB-LBMs [5,22] appear to be second-order schemes, all in the accuracy test through the simulation of Taylor-Green decaying vortices. However, some recent works reveal that the decaying vortices problem cannot serve as a reliable test case [23,24]. The reason is that analytical solution is artificially specified at the solid boundary, resulting in a continuous partial derivative across the interface which is too much idealistic for real applications [24]. In fact, it is usually not analytically tractable for Taylor-Green decaying vortices flow containing a physical solid boundary. Therefore, the decaying vortices should be substituted with other reasonable cases, such as the cylindrical Couette flow and the flow past a cylinder, in the accuracy test. In such situations, the IB-LBMs all degrade to be first-order accurate [6,23,53]. Many efforts are being devoted to improving the accuracy of IB-LBM. Most recently, Zhou et al. [25] reported a high-order Runge-Kutta based IB-LBM in 2014, which was claimed to improve the IB-LBM to be second-order accurate for the first time. The authors adopted the flow past a fixed/moving sphere in a tube as the accuracy test case (it is absolutely reliable), and a retraction distance (the solid boundary steps a little bit back to the geometry center) was used to improve the accuracy. They presented an optimal value of such distance for a sphere, i.e., 0.30 times of the spatial step. However, it is this retraction distance which helps to promote the accuracy of IB-LBM, because beyond a certain range, the proposed IB-LBM rapidly deteriorates to first-order accuracy. Since it is an open question to determine an optimal range of retraction distance applicable for different shapes of body, different discrete delta function and different flow configurations [63,64], there is still a need to develop a more general second-order accurate IB-LBM.

Other difficulty in the implementation of the conventional IB-LBM is to evaluate and incorporate a suitable external force within the framework of LBM. First, we should construct an external force model to compute the force. With the development of IBM, several force models have been proposed, including the spring force [26], feedback force [27], direct force [28] and momentum exchange force [29]. Note that the momentum exchange force scheme is a heuristic model derived directly from the bounce-back rule in the LBM, but has been demonstrated recently to be essentially a direct force model [6]. Those force models [26-29] have their own advantages and disadvantages, for example, the spring and feedback models are still stuck with the stiffness problem which requires a very small time step in computations [13]. Secondly, if the external force at the boundary has been evaluated and distributed to the ambient fluid (such distribution is usually accomplished with the help of discrete delta function), we should then determine the influence of this force in the framework of LBM because it is a kinetic method. There are many schemes for force-absorbing available in the literature. Guo et al. [8,30] systematically compared the performances of ten such methods. Cheng et al. [31] proposed another force-absorbing approach to incorporate the non-uniform and unsteady body force. Recently, Zheng et al. [32] presented a kinetic based force treatment. Note that the force-absorbing schemes should be corrected to adapt to different LBM models [33–35]. Furthermore, if the non-thermal and/or multi-phase flows encountered, it is still an open question to choose an appropriate force treatment in LBM [36–38].

Considering the difficulty and drawback outlined above for the conventional IB-LBM, in this work, a truly second-order IB-LBM for fluid-solid interactions and particulate flows is proposed. The present method improves the accuracy and preserves the flexibility concurrently of the traditional IB-LBM. Furthermore, it is based on the intrinsic nature of the LBM, i.e., directly correcting the distribution functions near the boundary to realize the no-slip boundary condition. Hence, there is no need for force evaluation and transformation in the present method. The remaining part of this paper is organized as follows. A methodology introduction is provided in Section 2. The numerical model is validated in Section 3. Finally, conclusions are summarized in Section 4.

## 2. Numerical methodology

The distribution function correction-based immersed boundary-lattice Boltzmann method is a combination of the IBM and LBM, for handling the boundary condition and the fluid flow, respectively. The LBM will be introduced briefly in this section, followed by the details of the immersed boundary scheme and the present improvements.

### 2.1. Lattice Boltzmann method

The lattice Boltzmann method is derived from the lattice gas automate (LGA), and has been greatly advanced to be an alternative incompressible flow solver in recent years [8,40]. As a kinetic scheme, a set of velocity distribution functions $f_i(\mathbf{x}, t)$ at position $\mathbf{x}$, time $t$ and discrete velocity $\mathbf{e}_i$, rather than the macroscopic quantities are tracked in LBM, which are subject to the lattice Boltzmann equation as

$$f_i(\mathbf{x}+\mathbf{e}_i\delta_t, t+\delta_t) - f_i(\mathbf{x}, t) = \Omega_i(f), \quad i = 0, 1, ..., b-1, \tag{1}$$

where $\Omega_i(f)$ denotes the discrete collision operator, $\delta_t$ the temporal step and $b$ the total number of discrete velocities. The MRT (multi-relaxation-time) collision model will be adopted instead of the widely-used LBGK (lattice Bhatnagar-Gross-Krook) model to avoid the unphysical numerical artifact and improve the stability, in which the collision term is expressed as [39]

$$\Omega_i(f) = -\sum_{j=1}^{b-1} \left(\mathbf{M}^{-1}\mathbf{S}\mathbf{M}\right)_{ij} \left(f_j - f_j^{eq}\right), \tag{2}$$

where $\mathbf{M}$ is a $b \times b$ transform matrix, and $\mathbf{S}$ is a relaxation matrix; $f_j^{eq}$ is the equilibrium distribution function which depends on the density $\rho$, velocity $\mathbf{u}$, and temperature $T$ of the gas and is typically defined as

$$f_j^{eq} = \omega_j \rho \left[1 + \frac{\mathbf{e}_j \cdot \mathbf{u}}{c_s^2} + \frac{(\mathbf{e}_j \cdot \mathbf{u})^2}{2c_s^4} - \frac{\mathbf{u}^2}{2c_s^2}\right], \tag{3}$$

where $\omega_j$ is the model-dependent weight coefficient, $c_s = \sqrt{RT}$ ($R$ is the gas constant) the lattice sound speed. For isothermal flows, $c_s$ is set to be $c/\sqrt{3}$ with $c = \delta_x/\delta_t$, where $\delta_x$ is the lattice spacing ($c = 1$ in this paper).

For simplicity and without loss of generality, the D2Q9 model (two dimensions with nine lattice velocities, $b = 9$) is employed for two-dimensional flows in the present study, where the velocity set and the corresponding weight coefficients are defined as

$$\mathbf{e}_i = \begin{cases} (0,0), \\ [\cos(i-1)\pi/4, \sin(i-1)\pi/4], \\ \sqrt{2}[\cos(i-1)\pi/4, \sin(i-1)\pi/4], \end{cases} \quad \omega_i = \begin{cases} 4/9, & i = 0, \\ 1/9, & i = 1, 2, 3, 4, \\ 1/36, & i = 5, 6, 7, 8. \end{cases} \tag{4}$$

The transform matrix $\mathbf{M}$ is given by [39]

$$\mathbf{M} = \begin{bmatrix} 1 & 1 & 1 & 1 & 1 & 1 & 1 & 1 & 1 \\ -4 & -1 & -1 & -1 & -1 & 2 & 2 & 2 & 2 \\ 4 & -2 & -2 & -2 & -2 & 1 & 1 & 1 & 1 \\ 0 & 1 & 0 & -1 & 0 & 1 & -1 & -1 & 1 \\ 0 & -2 & 0 & 2 & 0 & 1 & -1 & -1 & 1 \\ 0 & 0 & 1 & 0 & -1 & 1 & 1 & -1 & -1 \\ 0 & 0 & -2 & 0 & 2 & 1 & 1 & -1 & -1 \\ 0 & 1 & -1 & 1 & -1 & 0 & 0 & 0 & 0 \\ 0 & 0 & 0 & 0 & 0 & 1 & -1 & 1 & -1 \end{bmatrix} \tag{5}$$

and the relaxation matrix is [39]

$$\mathbf{S} = diag\left(s_\rho, s_e, s_\varepsilon, s_d, s_q, s_d, s_q, s_s, s_s\right) = diag\left(1/\tau_\rho, 1/\tau_e, 1/\tau_\varepsilon, 1/\tau_d, 1/\tau_q, 1/\tau_d, 1/\tau_q, 1/\tau_s, 1/\tau_s\right). \tag{6}$$

where $s$ and $\tau$ are the relaxation rate and relaxation time, and ($\tau_\rho$, $\tau_d$) and the remaining relaxation times are for conserved and non-conserved moments, which can take arbitrary values and should be set greater than 0.5, respectively.

Through Chapman-Enskog expansion, the macroscopic fluid density $\rho$ and velocity $\mathbf{u}$ can be derived respectively as the zeroth and first order moments of $f_i$,

$$\rho = \sum_{i=0}^{8} f_i, \quad \rho \boldsymbol{u} = \sum_{i=1}^{8} \boldsymbol{e}_i f_i. \tag{7}$$

The fluid pressure is directly obtained by $p = c_s^2 \rho$, and the fluid viscosity is related to the relaxation time $\tau_s$ for the shear moment as

$$\mu = \rho c_s^2 (\tau_s - 0.5) \delta_t. \tag{8}$$

### 2.2. Conventional immersed boundary scheme

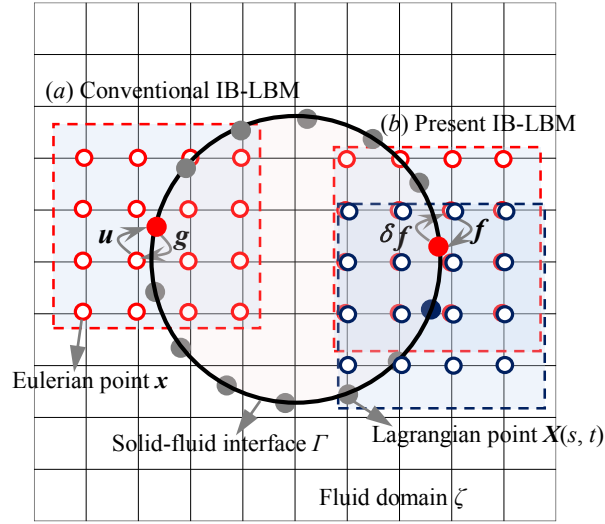

Figure 1. Schematic of the 2D immersed boundary method: (a) traditional scheme; (b) present scheme.

In IB-LBM, the IBM is usually used to impose the no-slip boundary condition at the solid-fluid interface $\Gamma$ (immersed in fluid $\zeta$), which is discretized into a set of Lagrangian markers $X(s, t)$, shown in Fig. 1. The main feature of the IBM is to model this interface as a generator of external body force; the force will then be distributed to the ambient Eulerian points $x$, appeared finally as a body force term in the governing equation [1,2]. In this manner, the no-slip boundary condition is replaced by a localized force field, and the fluid can sense the existence of the solid boundary. Generally, the implementation process of the IBM mainly contains the following three steps: evaluate the force, transfer the force and incorporate this force in the governing equation.

We should first calculate the local force at the Lagrangian point. There are several force schemes available in the literature. Lai et al. [26] proposed a spring force to mimic the rigid boundary as

$$\boldsymbol{G}(s,t) = k\left(\boldsymbol{X}^0(s,t) - \boldsymbol{X}(s,t)\right), \tag{9}$$

where $s$ is Lagrangian coordinate along the boundary, $X^0(s, t)$ denotes the expected position of the boundary, and $k$ is the coefficient of spring stiffness. In this force model, the Lagrangian points $X(s, t)$ are connected to their desired locations $X^0(s, t)$ with a spring. The positive spring constant $k$ is set to be sufficiently large. Therefore, the boundary points will be put back to their target positions once a small deviation emerges. The drawback for this spring force model is that a very small time step is required to obtain computational stability. Furthermore, it is a non-trivial task to determine a reasonable value of $k$. A feedback force model was reported by Saiki et al. [27], in which the boundary force is computed as

$$\boldsymbol{G}(s,t) = \alpha \int_0^t \left(\boldsymbol{u}(s,t) - \boldsymbol{U}(s,t)\right) dt + \beta \left(\boldsymbol{u}(s,t) - \boldsymbol{U}(s,t)\right), \tag{10}$$

where $u(s, t)$ is the fluid velocity and $U(s, t)$ is the real boundary velocity, both at position $X(s, t)$; $\alpha$ and $\beta$ are two negative and semi-empirical constants. The determination of $\alpha$ and $\beta$ is the core problem in this feedback force model, whose values are somewhat relevant to the temporal step. Taking reasonable values is to bring the velocity discrepancy between $u(s, t)$ and $U(s, t)$ to approach zero. Otherwise, it would seriously undermine the stability of the whole algorithm. Considering the limitations in the spring and feedback force models, Silva et al. [28] proposed a direct force scheme, in which the interfacial force is calculated without ad hoc constants as

$$G = \rho\left(\frac{\partial u}{\partial t} + u.\nabla u\right) + \nabla p - \mu\Delta u, \tag{11}$$

where the discrete form for the temporal term is $(U - u)/\delta_t$. It can be clearly observed that this force is directly obtained from the N-S equations, owing solid theoretical basis. Subsequently, that direct force model is further simplified to be a formulation of velocity deviation as [41]

$$G = \rho\frac{U - u}{\delta_t}. \tag{12}$$

Since the force model Eq. (12) has a very succinct form, no empirical parameters and no limitations for the temporal step, it is widely used in the field of IBM [54,55]. Furthermore, when combining this model with the LBM, some authors [42] claimed that the value of force should be doubled,

$$G = 2\rho\frac{U - u}{\delta_t}. \tag{13}$$

Considering that the no-slip boundary condition is gradually enforced in the IBM [21], Hu et al. [6] exhibited through rigorous mathematical deducing that the number 2 in Eq. (13) should be replaced by a relaxation parameter of $\lambda$ with an optimal range of 2.0–2.5. Recently, Niu et al. [29] made use of the bounce-back rule in the original LBM to construct a new force model as

$$G = \sum_{i=1}^{8} e_i\left(f_i - f_{\bar{i}}\right), \quad f_{\bar{i}} = f_i - 2\omega_i\rho\frac{e_i.U}{c_s^2}, \tag{14}$$

where $\bar{i}$ is the reverse direction of $i$, and $f_i$ is obtained from fluid by interpolation. However, it can be proved from Eq. (7) that this scheme is essentially equivalent to the direct force approach [6].

The force calculated by the above-mentioned models is then distributed to the ambient Eulerian points (Figure 1). In this step, we should determine the influence range and portion that the interfacial force imposes on the fluid. Many approaches can serve that propose for extrapolation, such as the radial basis function (RBF) [43], inverse distance weighted (IDW) scheme [44], and the discrete delta function [45]. The latter is more widely used in IBM, with a formulation of $D(x - X)$ as

$$D(x - X) = \frac{1}{\delta_x^2}\delta(x - x_l)\delta(y - y_l),$$

$$(a)\ \delta(r) = \begin{cases} 1 - |r|, & |r| \leq 1, \\ 0, & |r| > 1; \end{cases} \quad (b)\ \delta(r) = \begin{cases} \frac{1}{3}\left(1 + \sqrt{1 - 3r^2}\right), & |r| < 0.5, \\ \frac{1}{6}\left(5 - 3|r| - \sqrt{1 - 3(1 - |r|)^2}\right), & 0.5 \leq |r| < 1.5, \\ 0, & |r| \geq 1.5, \end{cases} \tag{15}$$

$$(c)\ \delta(r) = \begin{cases} \frac{1}{4}\left(1 + \cos\left(\frac{\pi|r|}{2}\right)\right), & |r| \leq 2, \\ 0, & |r| > 2; \end{cases}$$

where $(a)$, $(b)$ and $(c)$ are respectively the 2-, 3- and 4-point functions, and $x = (x, y)$ and $X = (x_l, y_l)$. The present work adopts the 4-point discrete delta function to spread the force to the Eulerian points,

$$g = \int_{\Gamma} GD(x - X)ds, \tag{16}$$

Note that this function is also used for the interpolation of fluid velocity $u$ at the Lagrangian point.

If the force exerted on the fluid points is obtained, finally we should add it in the governing equation of LBM, that is to say, the Eq. (1) as

$$f_i(x + e_i\delta_t, t + \delta_t) - f_i(x, t) = -\sum_{j=1}^{8}(M^{-1}SM)_{ij}(f_j - f_j^{eq}) + \delta_t M^{-1}(I - S/2)MF_i, \tag{17}$$

where $F_i$ is the force term that absorbs the external body force evaluated by the force models described above. Guo et al. [8,30] generalized ten specific forms of $F_i$, where two representative respectively are

$$F_i = n\frac{\rho \omega_i g \cdot e_i}{c_s^2}, \quad n = 1 \ or \ 0.5, \tag{18}$$

$$F_i = \omega_i \left[ \frac{e_i \cdot g}{c_s^2} + \frac{ug:(e_i e_i - c_s^2 I)}{c_s^4} \right]. \tag{19}$$

Note that when using the method in Eq. (19) to incorporate the force into the LBM, the fluid velocity should be calculated as

$$\rho u = \sum_{i=1}^{8} e_i f_i + 0.5\delta_t g. \tag{20}$$

Cheng et al. [31] proposed another type of $F_i$ as

$$M^{-1}(I - S/2)MF_i = \frac{1}{2}\left[ H_i(x + e_i\delta_t, t + \delta_t) + H_i(x, t) \right],$$
$$H_i = \omega_i \left\{ A + 3B \cdot \left[ (e_i - u) + 3(e_i \cdot u)e_i \right] \right\}, \tag{21}$$

where $A$ and $B$ are usually set to be 0 and $g$, respectively.

### *2.2.1. Iterative forcing approach*

The no-slip boundary condition is only approximately satisfied in the above-described IBM, because the forward/backward interpolation errors is inevitable to cause the unphysical streamline penetration [5,6,21]. To correct such deficiency, there are generally two strategies available in the literature: implicit velocity correction [5] and iterative forcing [22]. Considering that the former scheme might introduce extra computation loads [46], the idea of iterative forcing approach [22] will be adopted in the present study (even though actually there is no forcing in the present IB-LBM introduced in Section 2.3). The iterative forcing algorithm repeats the procedure of (i) velocity interpolation, (ii) obtain the force, and (iii) distribute the force multiple times, to gradually make the discrepancy of velocity at the Lagrangian point tend to be zero. It is found that within a finite number of iterations about 10, the residual error of such deviation will be reasonably small [6,22].

## 2.3. Distribution function correction-based immersed boundary scheme

As outlined above, there are difficulties existing in the implementation of the conventional IB-LBM, for instance, the model selections for force calculation and incorporating it into LBM. Furthermore, it is found in some recent studies that the traditional IB-LBM has only first-order accuracy in space [23,24]. Therefore, it motivates us to develop a second-order scheme with no forcing involved, that is, the distribution function correction-based IB-LBM.

The distribution function correction-based IB-LBM is derived from the original LBM, where the enforcement of boundary condition is accomplished by correcting the distribution functions at points near the boundary, as depicted in Fig. 2. In original LBM, the (simple or interpolated) bounce-back rule is usually adopted for such purpose. The reasonably obtained distribution function can make the no-slip boundary condition at the interface strictly enforced [61,62,65], since the macroscopic quantities are the moments of the distribution function, presented in Eq. (7). With those foundations in hand, hence, in the present IB-LBM, the aim is also to directly update the distribution functions around the boundary, both inside and outside (Fig. 1); so that if they are interpolated back at the interface, the no-slip boundary condition can be imposed through Eq. (7). For such propose, we first interpolate the uncorrected distribution function $(f_i(s,t))^*$ and its non-equilibrium part $(f_i^{neq}(s,t))^*$ at the Lagrangian point using the discrete delta function of Eq. (15),

$$(f_i(s,t))^{*,n} = \int_\zeta (f_i(x,t))^n D(x-X)dx, \quad i=0,1,...,8, \tag{22a}$$

$$(f_i^{neq}(s,t))^{*,n} = \int_\zeta (f_i^{neq}(x,t))^n D(x-X)dx, \tag{22b}$$

where $f_i^{neq} = f_i - f_i^{eq}$, and the superscript $n$ denotes the number of corrections (multiple corrections is needed, as explained next). The uncorrected non-equilibrium distribution function is approximated as the desired one at the boundary,

$$(f_i^{neq}(s,t))^n \cong (f_i^{neq}(s,t))^{*,n}. \tag{23}$$

Hence, plus the equilibrium part obtained by the information of the boundary (density $\rho$ and velocity $\boldsymbol{U}$), the desired distribution function at the boundary can be determined as

$$(f_i)^n = (f_i^{neq})^n + (f_i^{eq})^n = (f_i^{neq})^{*,n} + (f_i^{eq}(\rho,\boldsymbol{U}))^n. \tag{24}$$

A deviation is inevitable between the desired and interpolated distribution functions,

$$(\delta f_i(s,t))^n = (f_i)^n - (f_i)^{*,n}. \tag{25}$$

This discrepancy is then distributed to the ambient fluid points by the same discrete delta function,

$$(\delta f_i(x,t))^n = \int_\Gamma (\delta f_i(s,t))^n D(x-X)ds. \tag{26}$$

Finally, the distribution function at the Eulerian point can then be corrected one time,

$$(f_i(x,t))^{n+1} = (f_i(x,t))^n + (\delta f_i(x,t))^n, \tag{27}$$

Note that the above procedure, i.e., Eqs. (22) to (27) should be applied to all the Lagrangian points. A problem arises when we interpolate the revised distribution function back at the Lagrangian point, an un-negligible deviation between the revised and desired distribution functions could still be found, indicating that the no-slip boundary has not been well enforced according to Eq. (7). The reason is obvious because the range of the discrete delta function for the Lagrangian points may overlap with each other, as illustrated in Fig. 1. Therefore, it is insufficient to correct the distribution functions once. Here, inspired by the iterative forcing approach in the conventional IBM, we repeat the above-described process, i.e., obtaining the deviation of distribution function at the Lagrangian point and distributing it back to the Eulerian grids, Eqs. (22)–(27) multiple times. The residual error of deviation of distribution function is found to be reasonably small within a finite number of iterations $N$ about 10 ($n$ in Eqs. (22) and (27) takes 0, 1, … , $N$–1).

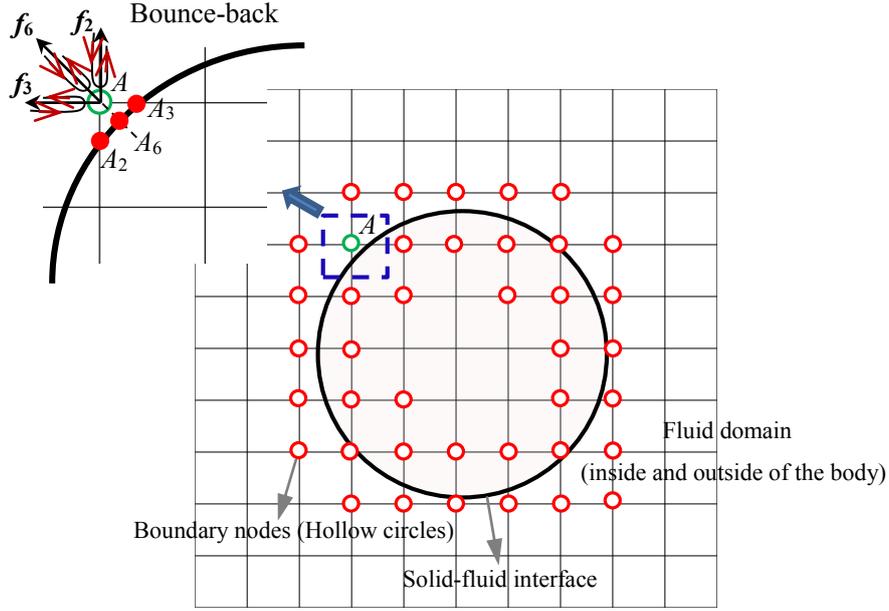

Figure 2. The boundary treatment in the original lattice Boltzmann method: correcting the distribution function at the boundary nodes (taking point *A* for example). The boundary nodes are those having at least one link intersected by the solid-fluid interface. In simple bounce back, the unknown distribution functions ($f_2$, $f_3$ and $f_6$) are equal to their post-collision counterparts pointing to the solid wall [65], that is, the intersection nodes ($A_2$, $A_3$ and $A_6$) are assumed to be located at the middle of two linked boundary nodes. Note that the unknown distribution functions can be obtained by some interpolated bounce back models for higher accuracy [61].

### 2.3.1. Procedure simplification and force evaluation

The implementation procedure of the present distribution function correction IB-LBM has been outlined above. To obtain the deviation of distribution function, both the distribution function and the non-equilibrium part are interpolated at the boundary point. In actual applications, such procedure can be further simplified as follows: we can just interpolate the equilibrium part and then calculate the deviation of equilibrium distribution function, in that the Eq. (25) can be rewritten as

$$\delta f_i = f_i - (f_i)^* = f_i^{eq} + f_i^{neq} - \left[ (f_i^{eq})^* + (f_i^{neq})^* \right] = f_i^{eq} - (f_i^{eq})^*. \tag{28}$$

Another significant issue in the application of IB-LBM is to evaluate the interaction force exerted on the body. In the preset IB-LBM, this hydrodynamic force $\boldsymbol{F}_s$ can be conveniently obtained as

$$\boldsymbol{F}_s = -\sum_n \sum_X \sum_i \boldsymbol{e}_i (\delta f_i)^n = -\sum_n \sum_X \sum_i \boldsymbol{e}_i \left( (f_i^{eq})^n - (f_i^{eq})^{*,n} \right), \tag{29}$$

where from right to left, the three summation operations are for the Eulerian points in the range of discrete delta function, the Lagrangian points at the boundary and the iterative numbers, respectively.

## 3. Results and discussions

The present IB-LBM is to be validated in the following two test cases: the cylindrical Couette flow and flow past a fixed cylinder. In the simulations, the relaxation times are set to be $\tau_s = 0.5 + \mu/(\rho c_s^2)\delta t$, $\tau_q = 1/1.8$, $\tau_e = 1/1.1$ and $\tau_\varepsilon = 1/1.25$ following Refs. [39,47]. We have also tested other set of relaxation times, and found that the values of $\tau_q$, $\tau_e$ and $\tau_\varepsilon$ have negligible influence on the present results.

## 3.1. Accuracy verification

### *3.1.1. Taylor-Green vortex with a cylinder*

The Taylor-Green vortex (TGV) flow containing an artificial circular cylinder is used first to test the accuracy order of the present IB-LBM. In this flow problem, many versions of IBMs have previously been considered to be second-order accuracy [5,6]. It is also noted that the reasonability of this flow taken as the accuracy validation case is questioned very recently [24].

The TGV flow with an artificial circular cylinder is defined in a $[-L, L] \times [-L, L]$ square box, subjected to the analytical solution as

$$
\begin{aligned}
u_x &= -u_0 \cos\left(\frac{\pi}{L}x\right)\sin\left(\frac{\pi}{L}y\right)e^{\left(-2\nu\left(\frac{\pi}{L}\right)^2 t\right)}, \\
u_y &= -u_0 \sin\left(\frac{\pi}{L}x\right)\cos\left(\frac{\pi}{L}y\right)e^{\left(-2\nu\left(\frac{\pi}{L}\right)^2 t\right)}, \\
\rho &= \rho_0 - \frac{u_0^2}{4}\left[\cos\left(\frac{\pi}{L}x\right)+\sin\left(\frac{\pi}{L}y\right)\right]e^{\left(-4\nu\left(\frac{\pi}{L}\right)^2 t\right)}.
\end{aligned}
\quad (30)
$$

The diameter of the cylinder is $L$, embedded in the center of the computation domain. In the simulation, the Reynolds number of the flow (Re = $u_0L/\nu$) is fixed at 10.0, and the relaxation time $\tau_s$ is set to be 0.65, as the same in Refs. [5,6,33,42]. The initial condition of the flow is obtained by Eq. (30) at $t = 0$. The time-dependent boundary conditions at the square and cylinder are given by Eq. (30), implemented by the non-equilibrium extrapolation scheme [62] and the present IB-LBM, respectively. Four sets of uniform grids with $L = 10, 20, 40$ and $80$ are used for simulation. At dimensionless time $t^* = tu_0/L = 1$, the numerical solution is obtained and the global error of the velocity field is quantified using the $L_2$-norm expressed as

$$
L_2 - \text{norm} = \sqrt{\sum_{i=1}^{N}\left[\left(u_x^c - u_x^a\right)^2 + \left(u_y^c - u_y^a\right)^2\right] \Big/ N}, \quad (31)
$$

where the superscripts $a$ and $c$ denotes the analytical and numerical solutions, respectively, and $N$ is the total number of Eulerian points in the flow field.

Velocity magnitude and the streamlines of the flow are presented in Fig. 3(a). Figure 3(b) plots the global $L_2$-norm error of the velocity, together with the data by the IB-LBM in literature [5,6,33,42] for comparison. Note that the Kang_1 and Kang_2 imply respectively the results using the 2-point and 4-point discrete delta functions in Kang et al. [42]. It can be clearly observed that second-order accuracy for the conventional IB-LBM is generally obtained by the previous works. However, the convergence rate of the present IB-LBM is almost in third-order for this TGV flow, seemingly in contrary to the documented conclusion that the original LBM is second-order accurate in space and hence the accuracy of hybrid scheme (IB-LBM) usually cannot exceed such value. This contradiction may result from the too much idealization of the TGV flow, since the flow macroscopic quantities and their derivations are assumed to be smooth at the inner circular interface. Therefore, this interface cannot be considered as the usual physical solid boundary that more concerned in applications. In general, for this idealized flow case, the present IB-LBM seems to have advantage over the conventional ones.

Cylinder embedded artificially

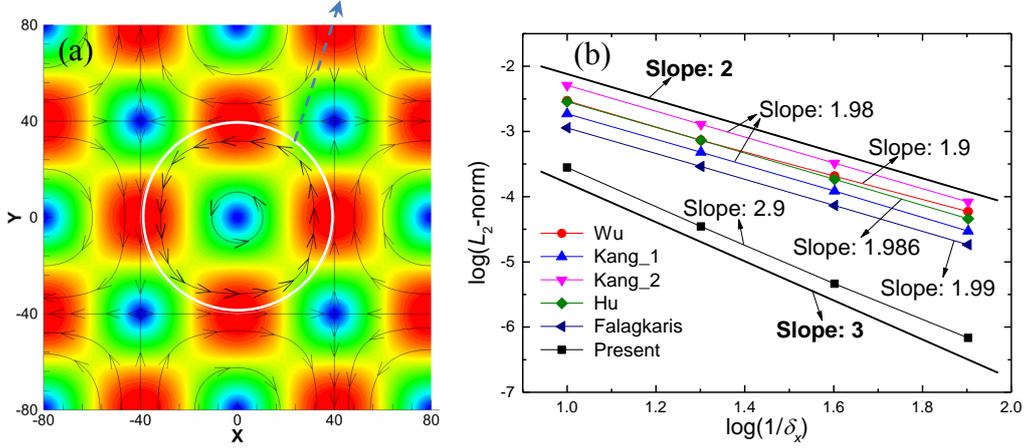

Figure 3. Contour of velocity magnitude and streamlines in the Taylor-Green vortex flow at $\delta_x = 1/80$ and $t^* = 1$.

### 3.1.2. Cylindrical Couette flow

Since the Taylor-Green vortex with an artificial cylinder essentially cannot served as a reasonable test case of accuracy outlined above, we further explore the accuracy of the present IB-LBM. It has been well-established that the LBM is a fully second-order scheme [8,40]. After combined with the conventional IBM, the accuracy reduces to be just first-order, which is found recently [23,24]. Therefore, it is important to verify the overall accuracy of the present IB-LBM. Note that here we choose the cylindrical Couette flow, other than the frequently-used decaying Taylor-Green vortex for conducting the numerical experiment, since the solid boundary exists physically in the former but is added artificially to the latter case. In the cylindrical Couette flow problem plotted in Fig. 4, the fluid is confined by the inner and outer rotating cylinders, with speeds $\omega_1$, $\omega_2$, and radii $R_1$, $R_2$, respectively. This flow is subjected to the N-S equations using the cylindrical polar coordinate as

$$\frac{d^2 u_\theta}{dr^2} + \frac{d}{dr}\left(\frac{u_\theta}{r}\right) = 0, \tag{32}$$

Coupled with the following boundary conditions ($\omega_2 = 0$),

$$u_\theta|_{r=R_1} = \omega_1 R_1, \quad u_\theta|_{r=R_2} = 0, \tag{33}$$

the steady solution of the flow can be obtained as

$$u_\theta = \frac{\left(R_1^2 - R^2 r^2\right)\omega_1}{\left(1 - R^2\right) r}, \quad u_r = 0, \tag{34}$$

where $u_\theta$, $u_r$ are the velocity components, $r$ is the radial distance and $R = R_1/R_2$ is the radius ratio.

In the simulations, it is set that the Reynolds number $Re = \rho U_1 (R_2 - R_1)/\mu = 10.0$ (where $U_1 = \omega_1 R_1$), $\tau_s = 0.65, 0.85$, and $R = 0.5, 0.8$. We calculate the $L_2$-norm to evaluate the numerical error as

$$L_2 - \text{norm} = \sqrt{\sum_{i=1}^{N}\left[\left(u_x - u_{\theta,x}\right)^2 + \left(u_y - u_{r,y}\right)^2\right] / N} \tag{35}$$

where $u_x$ and $u_y$ are the velocity components obtained by the present IB-LBM along the profile $\theta = \pi/2$ in the cylindrical polar coordinate. The velocity profile with $\delta_x = 1/40$ and $L_2$-norm error under different grid resolutions is presented in Fig. 5. The preset results agree well with those of analytical data (Fig. 4(a)). Furthermore, a generally second-order convergence rate can be observed for four sets of $\tau_s$ and $R$ (Fig. 4(b)). Hence, it can be concluded that the present IB-LBM is second-order accurate in space, preserving the accuracy of the original LBM.

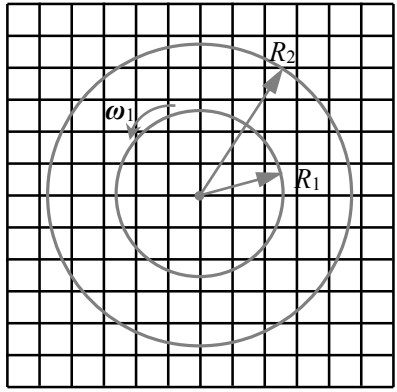

Figure 4. Schematic of the cylindrical Couette flow.

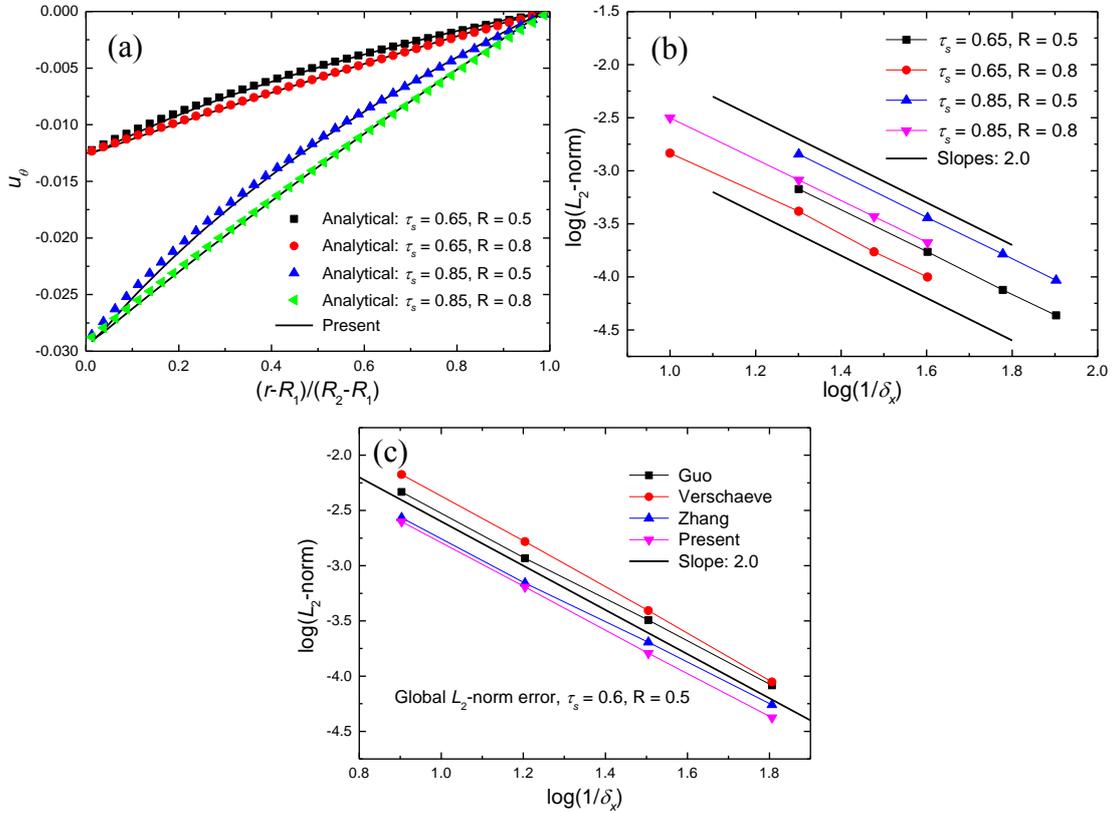

Figure 5. Velocity profile with $\delta_x = 1/40$ (*a*) and its relative norm error (*b*) at different relaxation times $\tau_s$ and radius ratios *R*. The global $L_2$-norm error is presented in (*c*).

## 3.2. Flow around a circular cylinder

The flow past a fixed circular cylinder is a classic fluid dynamic problem for which many numerical and experimental results are available in the literature [6,48–50]. This flow is controlled by the Re = $\rho_f U_0 D/\mu$, where $U_0$ and $D$ are the free stream velocity and cylinder diameter, respectively. For Re larger than 1 but lower than about 49, the long-time flow field is a steady state flow, and a recirculation region appears in the rear of the cylinder. Beyond that threshold, the flow could become unstable and eventually develops the Kármán vortex shedding. The quantitative results of this problem are usually the length of recirculation zone $L_w$, separation angle $\theta$, drag force $F_d$, lift force $F_l$ and frequency of vortex shedding $f$. The latter three are respectively defined in the non-dimensional form as

$$C_d = \frac{F_d}{0.5\rho_f U_0^2 D}, \quad C_l = \frac{F_l}{0.5\rho_f U_0^2 D}, \quad St = \frac{fD}{U_0}. \tag{36}$$

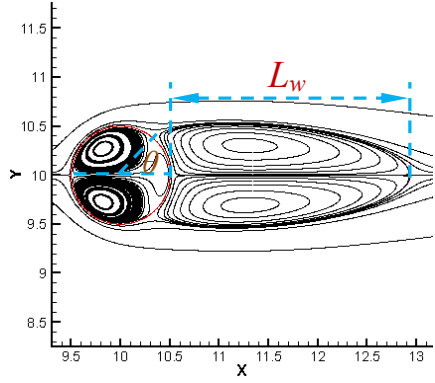

Figure 6. The streamlines around and inside the cylinder at Re = 40.

Table 1. Comparison of the results of flow over a circular cylinder at Re = 20, 40, 100 and 200.

| Re | 20 | | | 40 | | | 100 | | | 200 | | |
|---|---|---|---|---|---|---|---|---|---|---|---|---|
| | $C_d$ | $L_w/D$ | $\theta(°)$ | $C_d$ | $L_w/D$ | $\theta(°)$ | $C_d$ | $C_l$ | $St$ | $C_d$ | $C_l$ | $St$ |
| Russell[48] | 2.17 | 0.93 | 43.9 | 1.60 | 2.29 | 53.1 | 1.43 | 0.322 | 0.172 | 1.45 | 0.63 | 0.201 |
| Xu[49] | 2.23 | 0.92 | 44.2 | 1.66 | 2.21 | 53.5 | 1.423 | 0.34 | 0.171 | 1.42 | 0.66 | 0.202 |
| Linnick[50] | 2.16 | 0.93 | 43.9 | 1.54 | 2.28 | 53.6 | 1.38 | 0.337 | 0.169 | 1.37 | 0.7 | 0.199 |
| Hu[6] | 2.213 | 1.016 | — | 1.660 | 2.55 | — | 1.418 | 0.367 | 0.166 | 1.394 | 0.712 | 0.195 |
| Present | 2.181 | 0.987 | 44.2 | 1.643 | 2.46 | 53.2 | 1.415 | 0.358 | 0.166 | 1.405 | 0.719 | 0.196 |

In the simulations, size of the computational domain is (40$D$, 20$D$), and the circular cylinder is placed at the coordinate (10$D$, 10$D$). A uniform mesh system is adopted, with a grid resolution $\delta_x = D/40$. Constant velocity $U_0 = 0.1$ is specified at the inlet, and a free outflow is developed at the outlet. The inlet and outlet boundary conditions are both implemented by the non-equilibrium extrapolation method [62]. The streamlines near the circular cylinder are depicted in Fig. 6 for Re = 40, even though the inner fluid is fictitious in physics. It is clearly exhibited that no unphysical streamline-penetration appears, and the inner fluid cannot escape from the cylinder. Those results indicate that the no-slip boundary condition is accurately enforced in the present method. The recirculation length $L_w$, separation angle $\theta$, drag and lift coefficients ($C_d$ and $C_l$), and Strouhal number $St$ are presented in Table 1, including the data available in the literature [6,48–50] for comparison. Good agreement can be found between the present and the reference results.

### 3.3. Simulation of particulate flows

In this section, the flows with freely moving particles are to be simulated. The total force and torque exerted on the particle have to be calculated first for updating its position. The hydrodynamic force $\boldsymbol{F}_s$ coming from fluid is usually obtained by Eq. (29). However, when the particle is accelerated, term that account for the effect of inertial mass [54,55] should be added into the force and torque calculation as

$$F_s = -\sum_n \sum_X \sum_i e_i \left( \left(f_i^{eq}\right)^n - \left(f_i^{eq}\right)^{*,n} \right) + \rho_f V_s \frac{dU_s}{dt}, \quad (37)$$

$$T_s = -\sum_n \sum_X \sum_i e_i \left( \left(f_i^{eq}\right)^n - \left(f_i^{eq}\right)^{*,n} \right) \times (X_l - x_c) + \frac{\rho_f}{\rho_s} I_s \frac{d\phi_s}{dt}, \quad (38)$$

where $V_s$, $U_s$, $x_c$, $\rho_s$, $I_s$ and $\phi_s$ are the volume, velocity, mass center, density, momentum of inertia and angular velocity of the particle, respectively. Other forces that will need to be considered could include the force exerted on the particle due to the particle-particle or particle-wall collisions, which is usually given by the repulsive force scheme [4] as

$$F_c = \begin{cases} 0, & |x_i - x_j| > R_i + R_j + \xi, \\ \dfrac{c_{ij}}{\varepsilon_c} \left( \dfrac{|x_i - x_j| - R_i - R_j - \varsigma}{\xi} \right)^2 \left( \dfrac{x_i - x_j}{|x_i - x_j|} \right), & |x_i - x_j| \leq R_i + R_j + \xi, \end{cases} \quad (39)$$

where $c_{ij}$ is the force scale defined as the buoyancy force; $\varepsilon_c$ is the collision stiffness and takes 0.01 in the present study; $R_i$ and $R_j$ are the radii of the two particles centered at $x_i$ and $x_j$, respectively; $\xi$ is the threshold gap and set to be $0.05D$ ($D$ is the particle diameter). The Eq. (39) is also applied to the particle-wall collision, with $x_j$ being the position of a fictitious particle located symmetrically on other side of the wall with $R_j = R_i$. Note that the collision force always points to the particle center and hence does not make the particle rotate. After the force and torque exerted on the particle are obtained, the trajectory can then be tracked by the Newton's Second Law,

$$M_s \frac{du_s}{dt} = F_s + F_c - M_s \left(1 - \frac{\rho_f}{\rho_s}\right) g, \quad I_s \frac{d\phi_s}{dt} = T_s, \quad (40)$$

where $M_s$ is the particle mass. The translation and rotation velocities of the particle are updated by solving Eq. (40) with the first-order Euler method,

$$u_s^{n+1} = u_s^n + \delta_t (F_s + F_e)/M_s, \quad \phi_s^{n+1} = \phi_s^n + \delta_t T_s / I_s, \quad (41)$$

The particle position $x_s$ and rotation angle $\theta$ can then be obtained as

$$x_s^{n+1} = x_s^n + u_s^n \delta_t + \frac{1}{2}\delta_t^2 (F_s + F_e)/M_s, \quad \theta_s^{n+1} = \theta_s^n + \phi_s^n \delta_t + \frac{1}{2}\delta_t^2 T_s / I_s, \quad (42)$$

*3.3.1. A single particle settling in channel*

The first case is the sedimentation of a circular particle under gravity $|g| = 980$ cm$^2$/s in an open channel, which is well documented in numerical experiments [56–58]. As presented in Fig. 7, the width and height of channel are $W$ and $H$, respectively. A particle with diameter $D = 0.1$ cm falls freely in a fluid with density 1 g/cm$^3$ and kinematic viscosity 0.01 cm$^2$/s. In the simulations, $W$ is equal to $4D$ and $H$ is set to be 400/13 $D$; the particle is initially placed at a distance of $0.31W$ from the channel centerline and resolved by 26 lattice units along the diameter; the relaxation time $\tau_s$ takes a value of 0.6; four sets of particle-fluid density ratios, $\rho_s/\rho_f = 1.0015, 1.003, 1.01$ and $1.03$ are considered. The computational parameters are the same as those in Refs. [57,58]. The fluid velocity at the inlet of channel is assumed to be zero, and the free-stream boundary condition is applied at the channel outlet.

The time histories of the particle trajectory and translational velocity are presented in Fig. 8. The results reported by Hu et al. [56] using the arbitrary-Lagrangian-Eulerian (ALE) method are also included for comparison. From the Fig. 8(a), we can see that the particle will gradually move to the centerline of the channel for all density ratios $\rho_s/\rho_f$. However, at larger $\rho_s/\rho_f$ (1.01 and 1.03), the particle can oscillate before approaching the final equilibrium position. That conclusion is confirmed in Fig. 8(b), where the value of velocity in the width direction is found to fluctuate around zero. Furthermore, it is clearly observed that the present results agree well with those in Hu et al. [56].

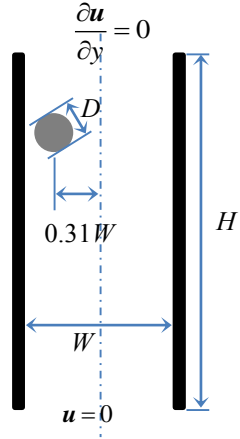

Figure 7. Schematic of a particle settling in a channel.

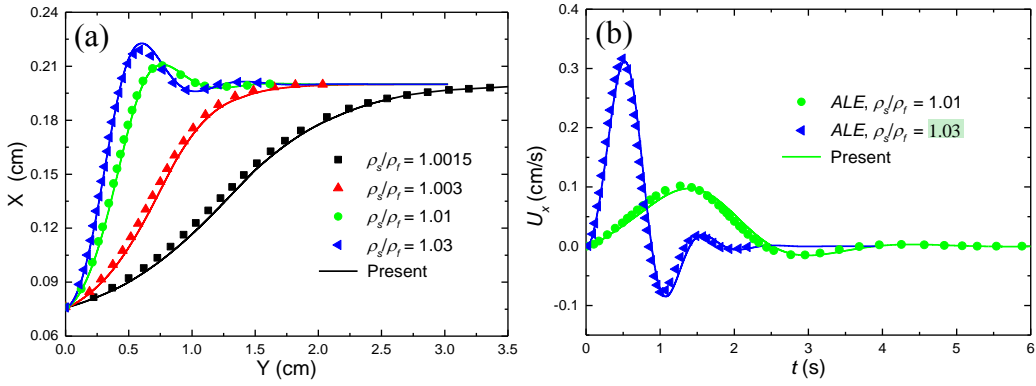

Figure 8. Time histories of the particle trajectory and translational and rotational velocities. Lines; Present; Scatters: ALE method [56].

### *3.3.2. DKT of two particles in channel*

In many application scenarios, a particle can undergo frequent inter-particle interactions, apart from interacting with the fluid. In this subsection, the simulation of a particle pair settling under gravity is performed to further evaluate the present IB-LBM in modeling multiple particle systems. This standard test case has been performed in many previous studies. The size of the channel is 2 cm × 8cm with $D$ = 0.2 cm being the particle diameter, and the kinematic viscosity of fluid is $\nu$ = 0.01 g/(cm. s). The upper and lower particles are identical with the same density of $\rho_p$ = 1.01 g/cm$^3$. The initial positions of the two particles are (0.999 cm, 7.2 cm) and (1.0 cm, 6.8 cm), respectively. In order to break the strong symmetry of the flow and to induce tumbling motion later on, a slight deviation in the *x*-direction is set intentionally for the first particle. In the simulations, the particle is resolved by 20 grids corresponding to a uniform mesh with size of 200 × 800 for the computational domain, and the relaxation time $\tau_s$ is set to be 0.8. The parameters in the present study are chosen the same as those in [4,29,59,60].

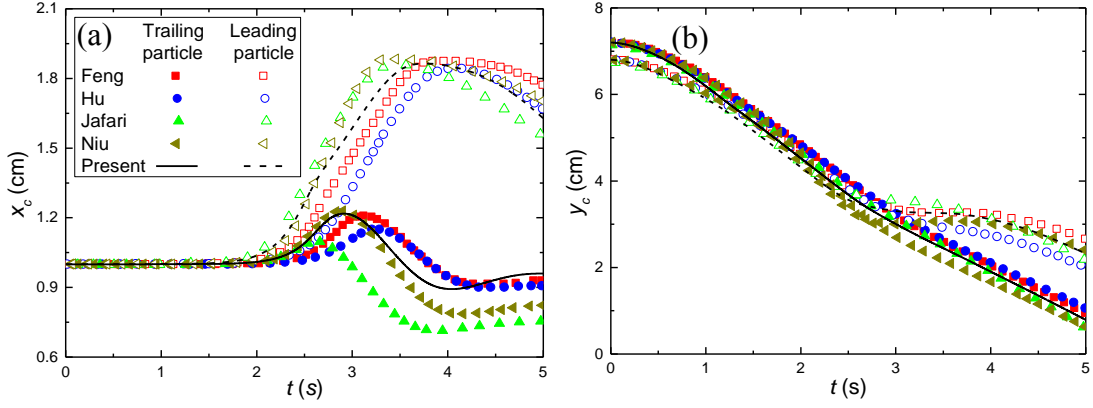

Figure 9. The trajectory of the two particles in the *x* (a) and *y* (b) directions.

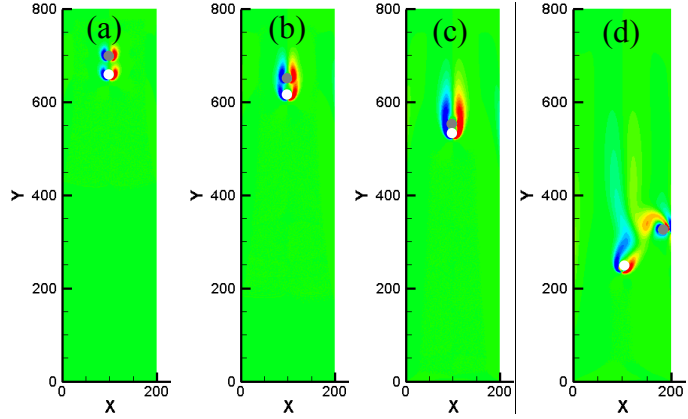

Figure 10. The instantaneous vorticity at $t = 0.4$ (*a*), 0.8 (*b*), 1.4 (*c*), 3.5 s (*d*) during the settling process of two particles in a channel.

Figure 9 presents the instantaneous positions of the two particles during the settling process. The DKT (drafting, kissing and tumbling) phenomenon is clearly reproduced. Initially, the two particles are located along the centerline of channel with a relatively small gap. After being released from rest in the still fluid, both particles begin to descend under gravity, as shown in Fig. 10(*a*). While the leading particle is falling down, it creates a wake with lower pressure. As the trailing particle comes close to the leading one, it is drafted into the wake and experiences a much smaller drag (Fig. 10 (*b*)). Hence, the trailing particle moves faster than the leading one, and eventually catches up, and then kisses and impels the latter. This stage persists for some time, during which the particles form a doublet and fall downwards together (Fig. 10 (*c*)). However, that state is unstable as indicated in [29,60], because of some symmetry breakings such as the fluctuating wake. As a result, the sedimentation process turns into the tumbling stage, where the particles start to separate from each other (Fig. 10 (*d*)). The time history of the center to center distance of the particles is given in Fig. 11. It can be observed that after about 0.514 seconds, the gap decreases gradually. At about $t = 1.326$ s, the distance approaches to a local minimum value, indicating a contact with each other, and this kissing stage lasts about 1.166 seconds. Finally, at about $t = 2.492$s, the distance increases and the particles start to separate from each other. As shown in Fig. 8, the DKT processes predicted by the IB-LBM agree well with those reported by Hu et al. [59] using the LBM. However, noticeable differences in the tumbling stages are seen for the two results. As indicated in [59,60], the difference can be expected in that the dynamics in the tumbling phase relies heavily on the growth rate of the numerical uncertainties and the boundary treatments as well as the collision models. Hence, the present IB-LBM can be generally considered to be able to produce reasonable results for the DKT dynamics of two particles.

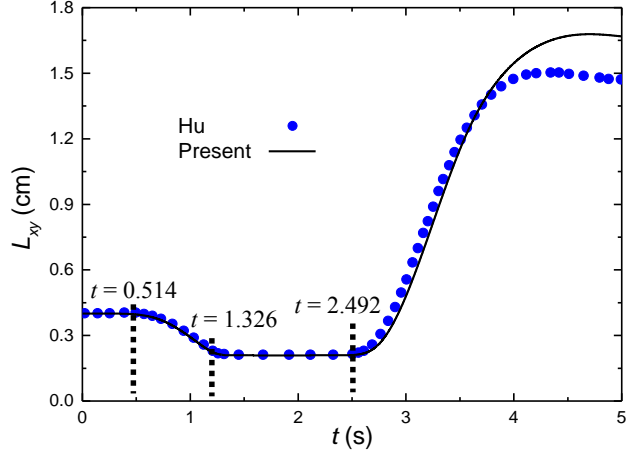

Figure 11. Time history of the distance between the two particles.

## 4. Conclusions

A simple and efficient distribution function correction-based IB-LBM is proposed in this paper. The improvement of the present IB-LBM is that it promotes the accuracy of the conventional ones from first- to second-order, while preserving the flexibility of this hybrid method. In addition, there is no need for present scheme to construct the models for evaluating and distributing the force density at the solid boundary. Instead, derived from the original LBM, the boundary treatment is also to obtain the correct distribution functions near the boundary in the present IB-LBM. Second-order accuracy of this method is reasonably confirmed in the simulation of cylindrical Couette flow. The present method is validated in the flow past a fixed cylinder. No streamline penetration is found, indicating the exactly enforcement of the no-slip boundary condition. Furthermore, in the simulations of one and two particles settling under gravity, good agreements can be observed comparing with the data available in literature.

The present study proposes the basic ingredient of the IB-LBM based on the distribution function correction, and the simulations are limited to the two-dimensional flow problems. It is straightforward to extend the present IB-LBM to three-dimensional flows, by correcting the distribution function using the three-dimensional discrete delta functions. The application to the non-isothermal flows will also be studied in future work.

### Acknowledgments

S. Tao is very much thankful to Prof. Guo and Prof. Wang for insightful discussions. This work was supported by the National Natural Science Foundation of China (Grant Nos. 51406036 and 51306038).### References

[1] Peskin, C. S. (2002). The immersed boundary method. *Acta numerica*, *11*, 479–517.
[2] Uhlmann, M. (2005). An immersed boundary method with direct forcing for the simulation of particulate flows. *Journal of Computational Physics*, *209*(2), 448–476.
[3] Akiki, G., & Balachandar, S. (2016). Immersed boundary method with non-uniform distribution of Lagrangian17